# Stable and Solution-Processable Cumulenic sp-Carbon Wires: A New Paradigm for Organic Electronics

*Stefano Pecorario, Alberto D. Scaccabarozzi, Daniele Fazzi, Edgar Gutiérrez-Fernández, Vito Vurro, Lorenzo Maserati, Mengting Jiang, Tommaso Losi, Bozheng Sun, Rik R. Tykwinski, Carlo S. Casari, and Mario Caironi*

Solution-processed, large-area, and flexible electronics largely relies on the excellent electronic properties of sp²-hybridized carbon molecules, either in the form of π-conjugated small molecules and polymers or graphene and carbon nanotubes. Carbon with sp-hybridization, the foundation of the elusive allotrope carbyne, offers vast opportunities for functionalized molecules in the form of linear carbon atomic wires (CAWs), with intriguing and even superior predicted electronic properties. While CAWs represent a vibrant field of research, to date, they have only been applied sparingly to molecular devices. The recent observation of the field-effect in microcrystalline cumulenes suggests their potential applications in solution-processed thin-film transistors but concerns surrounding the stability and electronic performance have precluded developments in this direction. In the present study, ideal field-effect characteristics are demonstrated for solution-processed thin films of tetraphenyl[3]cumulene, the shortest semiconducting CAW. Films are deposited through a scalable, large-area, meniscus-coating technique, providing transistors with hole mobilities in excess of 0.1 cm² V⁻¹ s⁻¹, as well as promising operational stability under dark conditions. These results offer a solid foundation for the exploitation of a vast class of molecular semiconductors for organic electronics based on sp-hybridized carbon systems and create a previously unexplored paradigm.

## 1. Introduction

The synthesis of novel organic semiconductors, along with advances in processing techniques and device engineering, has boosted the field of large-area flexible and printed electronics. These advances have enabled a plethora of applications such as organic light-emitting diodes,[1,2] organic photovoltaics,[3,4] organic thermoelectrics,[5,6] organic field-effect transistors (OFETs),[7–10] organic (bio)sensors,[11–13] and neuromorphic devices.[14,15] In this context, organic field-effect transistors (OFETs) are not only relevant for their direct technological application, but they also represent an ideal test-bed to investigate thin-film electrical properties. Organic semiconductors are typically classified in two main families, namely conjugated polymers and small molecules. The former, polymers, are particularly appealing as a result of their solution processability, and OFETs with charge mobility above the standard for hydrogenated amorphous silicon (0.5–1 cm² V⁻¹ s⁻¹) have been extensively reported.[16] The latter, small molecules, are prone to arrange in ordered molecular crystals, and through several years of chemical tailoring and fine tuning of the films processing, small-molecule OFETs with field-effect mobility >10 cm² V⁻¹ s⁻¹ have been achieved.[17–19] The chemical root of the π-conjugation of these materials is associated with the sp²-hybridization of carbon atoms in their backbone. This peculiar trait is also common to

S. Pecorario, A. D. Scaccabarozzi, V. Vurro, L. Maserati, M. Jiang, T. Losi, M. Caironi
Center for Nano Science and Technology@PoliMi
Istituto Italiano di Tecnologia
via Giovanni Pascoli 70/3, Milano 20133, Italy
E-mail: mario.caironi@iit.it

The ORCID identification number(s) for the author(s) of this article can be found under https://doi.org/10.1002/adma.202110468.

S. Pecorario, C. S. Casari
Department of Energy
Micro and Nanostructured Materials Laboratory – NanoLab
Politecnico di Milano
Via Ponzio 34/3, Milano 20133, Italy

D. Fazzi
Department of Chemistry "Giacomo Ciamician"
Università di Bologna
Via F. Selmi, 2, Bologna 40126, Italy

E. Gutiérrez-Fernández
POLYMAT
University of the Basque Country
UPV/EHU, Av. de Tolosa 72, San Sebastián 20018, Spain

B. Sun, R. R. Tykwinski
Department of Chemistry
University of Alberta
Edmonton, AB T6G 2G2, Canada

low-dimensional carbon nanostructures that show exceptional (opto-)electronic properties, such as graphene, fullerenes, and carbon nanotubes.[20–23]

There exists an alternative class of π-conjugated molecules with backbones formed from sp-hybridized carbon atoms. Known as carbon atom wires (CAWs) or linear carbon chains, these molecules are regarded as an oligomeric form of the so-called "carbyne," a purely one-dimensional carbon allotrope made of an infinite chain of sp-carbons.[24] Several theoretical and computational studies have predicted remarkable properties for carbyne, such as an exceptional Young's modulus[25] and a high surface area,[26] as well as significant optical absorption,[27] thermal conductivity,[28] nonlinear optical properties,[29–32] and charge mobilities, which might rival those of graphene and carbon nanotubes.[33] For decades, chemists have been attempting to overcome the chemical reactivity of molecules composed of sp-carbon, developing syntheses to provide longer and longer CAWs in the ultimate pursuit of the "carbyne" limit.[34–37] A successful approach to stable CAWs in powder form consists of endcapping the sp-carbon chain with bulky terminal groups that prevent chain cross-linking through steric hindrance. Endgroups are also exploited to control the bond-length alternation (BLA) of the CAWs that, in turn, affects the vibrational, optical, and electronic properties.[24,38–42] Indeed, two isomeric structures of CAWs are possible based on BLA: cumulenes, displaying a sequence of quasi-double bonds (BLA < 0.1 Å) and oligoynes with a sequence of alternated single and triple bonds (BLA > 0.1 Å). The structure of cumulenes enables extensive π-electron delocalization along the backbone, resulting in a lowering of the HOMO-LUMO gap. On the contrary, the bond alternated structure of oligoynes is characterized by localization of π-electrons on the triple bonds, which accounts for a larger HOMO-LUMO gap. While recently long oligoynes with a record length of 48 sp-carbons have been successfully isolated,[37] the synthesis and handling of long cumulenes is more problematic due to a drastically increased reactivity with the chain length, and [9]cumulenes (i.e., CAWs with 9 consecutive double bonds) are the longest derivatives reported so far.[43] Despite challenges posed by chemical instability, there is a growing interest in exploiting CAWs in molecular electronics and optoelectronics,[44] justified by intriguing evidence such as increased conductance with longer cumulene length.[45–48]

Surprisingly, the properties of oligoynes and cumulenes in thin films toward large-area electronics remain almost completely unknown.[44,49,50] Recently, the first OFETs based on tetraphenyl[3]cumulene (also known as tetraphenylbutatriene and with the abbreviated form [3]Ph, which will be used hereafter) have been demonstrated. A simple drop-casting procedure gave micrometers-long, needle-like molecular crystals of [3]Ph.[51] Being the shortest of the cumulenes series, [3]Ph is constructed of three cumulated carbon-carbon double bonds (BLA = 0.088 Å) and it is terminated at each end by two phenyl groups (**Figure 1**a). This short cumulene was selected as a model material because it can be synthesized on a gram scale, it is soluble in organic solvents, thermally stable up to 250 °C, and characterized by an optical gap in the visible range (properties are summarized in Table S1 in the Supporting Information). A random network of [3]Ph microcrystals shows clear p-type semiconducting behavior, with a field effect mobility conservatively estimated to be $\approx 2 \times 10^{-3}$ cm$^2$ V$^{-1}$ s$^{-1}$.[51] The needle-like texture, despite exhibiting a remarkable crystallinity, presents intrinsic limits associated with poor coverage of the active area, inadequate carrier injection, and the presence of hysteresis in the transfer curve characteristics. Furthermore, the stability of [3]Ph-based OFETs in operational conditions represents a natural challenge that needs to be addressed. Therefore, serious questions and concerns remain for the credible adoption of sp-hybridized molecules for solution-processed thin-film electronics.

Herein, we provide proof that transistors can be based on solution-processed polycrystalline cumulenic thin films, with charge carrier mobilities close to that of a:Si and promising stability in dark-conditions. In particular, we present OFETs based on thin films of [3]Ph deposited from solution via a scalable, large-area wire-bar coating technique, and we analyze the impact of the deposition temperature on morphological, optical, structural and electrical properties of the films by means of extensive structural and electrical characterization. The optimized deposition temperature for [3]Ph (i.e., 80 °C) provides a higher degree of crystallinity, as well as fewer defects. The resulting devices show field-effect mobility values exceeding 0.1 cm$^2$ V$^{-1}$ s$^{-1}$, in both the linear and saturation regime, marking an improvement of two orders of magnitude relatively to previous studies on sparse microcrystalline networks.[51] Supported by density functional theory (DFT)/time-dependent DFT (TD-DFT) calculations, we emphasize that the measured mobilities represent a lower bound to the intrinsic mobility of [3]Ph as the result of marked charge transport anisotropy, the lack of preferential orientation in the plane of the films, and the presence of grain boundaries in polycrystalline films. Finally, transistors based on [3]Ph show promising ambient stability under dark conditions, therefore addressing a major concern for the effective application of sp-based semiconducting molecules in electronic devices. These findings establish cumulenic sp-carbon atom wires as promising class of organic semiconductors for large-area, printed electronics.

## 2. Results

### 2.1. Deposition and Characterization of Thin Films

To investigate the potential of cumulenes toward large-area printed electronics, we began by optimizing the deposition of thin films of [3]Ph from solution (10 g L$^{-1}$ in 1,2-dichlorobenzene) employing wire-bar coating, which is a scalable bar-assisted meniscus-shearing technique (Figure 1b).[52–54] The deposition temperature has a strong impact on the crystallization of solution-processed small-molecule organic semiconductors, which in turn defines the charge transport characteristics in the solid state.[55,56] Accordingly, we systematically varied the deposition temperature using a heating bed and examined the impact on the optical, morphological, structural, and electrical properties of the thin films. The choice of a high-boiling-point solvent such as 1,2-dichlorobenzene led to homogeneous thin films with excellent coverage over an area of 2 cm$^2$ and a wide range of deposition temperatures, spanning from 50 to 110 °C.

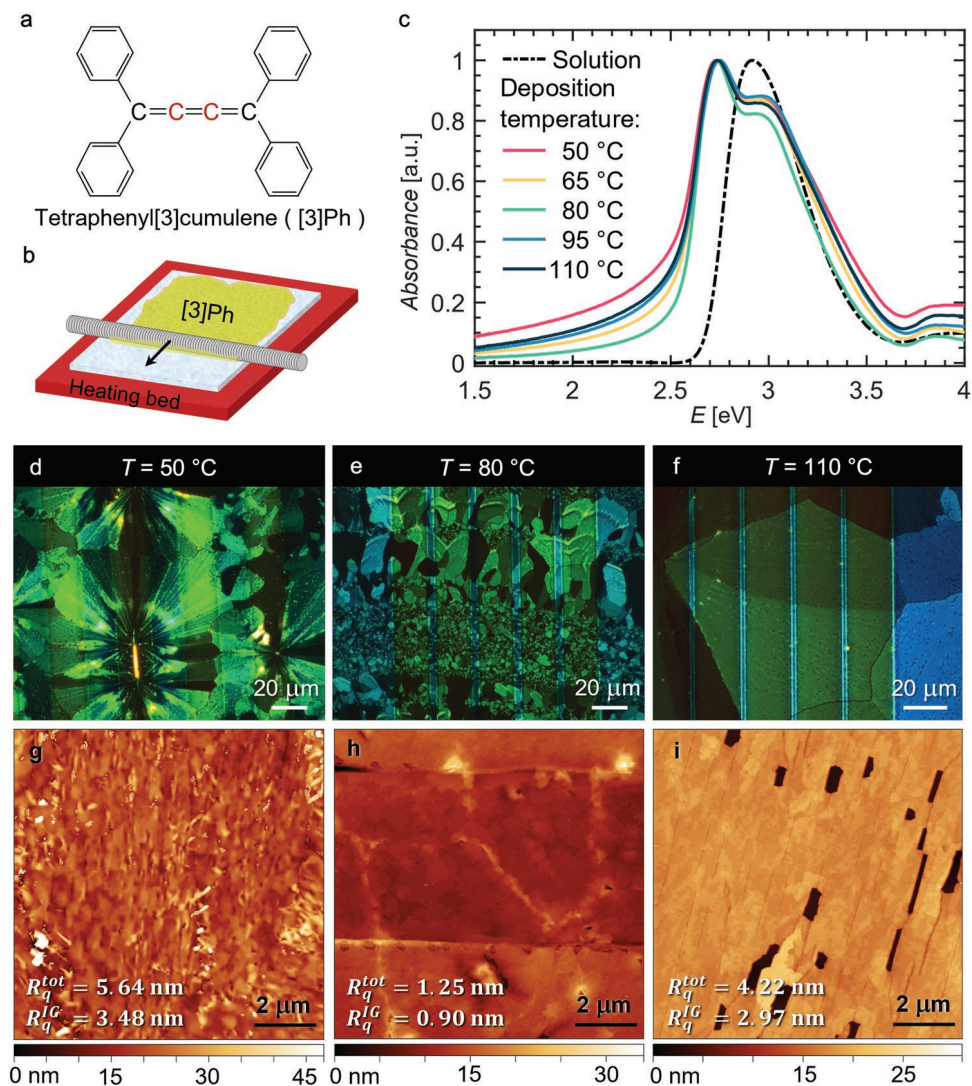

**Figure 1.** a) Molecular structure of [3]Ph. The sp-hybridized carbons constituting the cumulenic backbone are indicated in red. b) Scheme of the deposition of large-area [3]Ph thin films from solution (10 g L$^{-1}$ in 1,2-dichlorobenzene) on a glass substrate via wire-bar coating. c) Normalized absorption spectra of [3]Ph in solution (10$^{-4}$ M in 1,2-dichlorobenzene, dashed black line) and of thin films deposited at different temperatures (full colored lines). d–f) POM images of films of [3]Ph bar coated on top of gold interdigitated source and drain electrodes at 50, 80, and 110 °C, respectively. g–i) AFM topographies of films of [3]Ph deposited at 50, 80, and 110 °C, respectively. The derived root-mean-square roughness values are reported, distinguished in intragrain roughness ($R_q^{IG}$) and total roughness ($R_q^{tot}$).

To assess aggregation in thin films, we compared the UV–vis absorption spectra of [3]Ph in a dilute solution (Figure 1c, black dotted line) and in films deposited at different temperatures (Figure 1c, colored solid lines). In solution, only one intense absorption peak at 2.92 eV characterizes the low energy region of the spectrum. For thin films, on the other hand, two absorption bands are observed at 2.73 eV and 2.97 eV, with the former significantly red-shifted and the latter slightly blue-shifted with respect to the absorption band in solution (Figure S1, Supporting Information). The intense peak at 2.73 eV has been attributed to the occurrence of intermolecular interactions of [3]Ph in the solid state.[51,57] All of the absorption spectra of thin films have bands located at the analogous energies, suggesting a common intermolecular arrangement independent of the deposition temperature.

From a morphological point of view, the films of [3]Ph are polycrystalline and display a clear variation in the crystalline grain dimensions depending on the deposition temperature, which can be appreciated by the polarized optical microscopy (POM) images of films deposited at 50, 80, and 110 °C (Figure 1d–f and Figure S2, Supporting Information). Based on mechanical profilometry and atomic force microscopy (AFM) analyses (Figure 1g–i), we determined that the film thickness varies, on average, between 15 and 30 nm, with a slight thinning of films deposited at higher temperatures. The films deposited at 50 °C present spherulitic-like structures, measuring tens of micrometers, which grow radially from a central nucleation site and produce the characteristic Maltese crosses (Figure 1d). In addition, needle-like crystals are visible on the surface, resembling on a smaller scale the long crystals formed by drop-casting (Figure S3, Supporting Information).

The films deposited at 80 °C appear more homogeneous and feature few micrometer-scale 2D crystalline grains and no needle-like crystals (Figure 1e). The increased homogeneity and smaller grain size can be rationalized in terms of faster solvent evaporation and reduced time for nucleation and crystal growth.[58–60] Further increasing the temperature to 110 °C leads to films with qualitatively similar 2D grains, although the grain size is increased up to several tens of micrometers (Figure 1f). The latter is likely associated with the higher deposition temperature, which can lead to an increased rate of crystallization, according to the bell-shaped nature of the crystallization rate as a function of temperature.[61,62]

The surface morphology of the films was further investigated by AFM. The spherulitic structures, typical of films deposited at 50 °C, present a rough surface, predominantly due to the small needle-like crystals that abound close to the grain boundaries (Figure 1g and Figure S4, Supporting Information). The 2D crystalline grains obtained at 80 °C have a much smoother surface, and it is even possible to identify terraces with regular steps with a height of about 1 nm (Figure S5, Supporting Information), which is compatible with the thickness of a single monolayer of [3]Ph molecules stacked with the sp-chain parallel to the substrate (further details provided with crystalline structure analysis, vide infra). Finally, the larger grains of the films deposited at 110 °C include lines of defects and pits, as evident in Figure 1i, which relates to a topographic image acquired within the same grain (see also Figure S6a,b, Supporting Information). These defects might originate from the cracking of the films due to the rapid cooling from 110 °C to ambient temperature. We derived the root-mean-square roughness values of the films ($R_q$) and distinguished between the intragrain roughness ($R_q^{IG}$), considering only the portion inside a single-crystalline grain, and the total one ($R_q^{tot}$), which includes more crystalline domains and the grain boundaries. A detailed illustration of the roughness extraction procedure is provided in Figures S4–S6 in the Supporting Information. Importantly, the small 2D grains distinctive of the deposition at 80 °C have a total roughness that is even lower than the intragrain roughness of the films deposited at 50 and 110 °C.

To investigate the structural arrangement of the [3]Ph molecules in thin films, we performed grazing-incidence wide-angle X-ray scattering (GIWAXS) measurements on films obtained at the various deposition temperatures (**Figure 2**a–c, Supporting Information). The presence of bright, point-like diffraction peaks corroborates crystallinity and anisotropy of the films with respect to the substrate. The assignment of Miller indices is provided in Figure S7 in the Supporting Information. The unit cell is reported to be triclinic and formed by a basis of two [3]Ph molecules closely packed as illustrated in the 2 × 2 supercell representation of Figure 2d,e.[63] Regardless of deposition temperature, the diffraction patterns of the films fit well with the single-crystal structural analysis,[51,63] meaning that the crystal structure of the single crystal is retained in the polycrystalline thin films. From the angular position of the peaks in the diffractograms (in-plane/out-of-plane diffractions reported in Figure S8 in the Supporting Information), it is possible to determine the orientation of the lattice axis. In this regard, the location of the (001) reflection in the out-of-plane direction ($Q_Z$ = 6.22 nm$^{-1}$) are associated with a $d$-spacing of 1.01 nm and indicates that the molecules are stacked along the normal direction of the substrate plane and separated by the phenyl groups. At the same time, the reflections (200) at $Q$ = 12.8 nm$^{-1}$ and (120) at $Q$ = 16 nm$^{-1}$, are associated with shorter packing distances located near the $Q_R$ axis. These reflections correspond to the diffraction along the sp-chains π-stacking and indicate that the cumulene π-stacking occurs parallel to the substrate (Figure 2d,e). Furthermore, the presence of Bragg rods confirms that the *a* and *b* vectors of all grains lie in the substrate plane, while the azimuthal orientation is random.[64] This arrangement is consistent with the polycrystalline nature of the films, as previously observed based on the POM images. Interestingly, the computed unit cell based on the GIWAXS pattern of thin films deposited at 110 °C shows a contraction along the *a*-axis and an elongation along the *b*-axis, with respect to films cast at lower temperatures (Figure 2f, details in Figure S9 and Table S2 in the Supporting Information). Instead, the molecular distance along the *c*-axis is not affected by the deposition temperature.

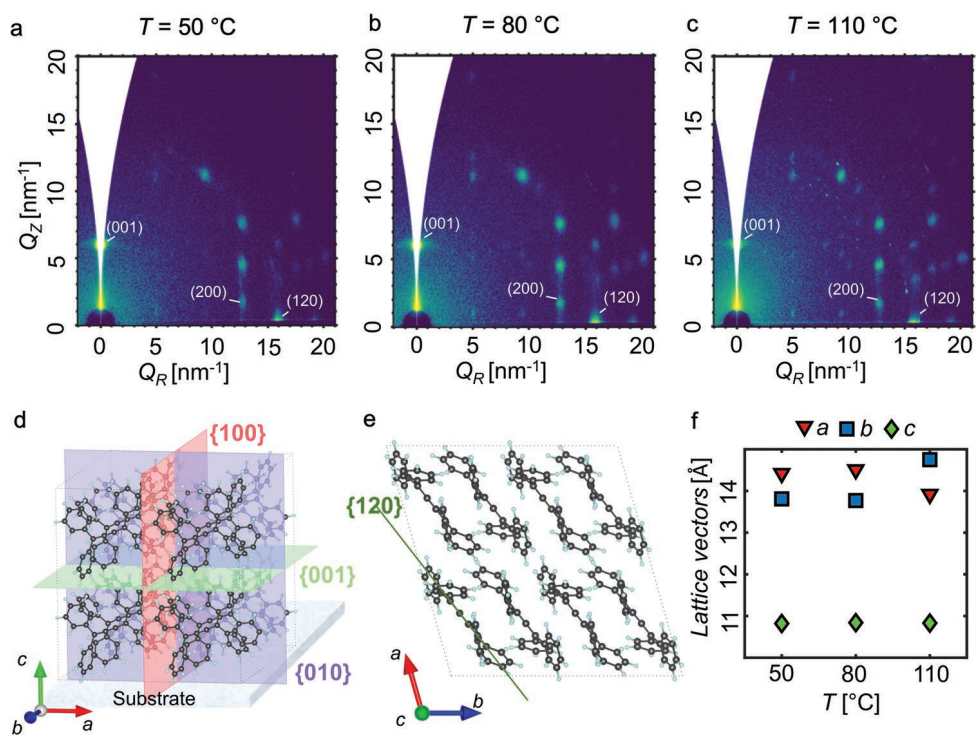

**Figure 2.** a–c) GIWAXS patterns of thin films of [3]Ph deposited at 50, 80, and 110 °C, respectively. d) Graphical representation of the crystal orientation of [3]Ph (2 × 2 supercell) with respect to the substrate; molecules are aligned with the cumulenic backbone quasi-parallel to the substrate. e) Portrayal of the in-plane lamellar molecular stacking of [3]Ph. f) Length of the lattice vectors as a function of the deposition temperature.

## 2.2. Excitonic and Electronic Molecular Couplings

DFT/TD-DFT calculations on a single molecule and on molecular dimers were performed to correlate the excited-state properties with the molecular packing and to provide insight into the optical absorption changes upon film deposition (Figure 1c). The low-energy computed dipole active excited state for a single molecule of [3]Ph is 3.13 eV. Exciton splitting occurs upon aggregation (i.e., passing from the single molecule to the dimer model), thus resulting in a higher and a lower energy state.[65] In the dimer-based analysis, we considered all nearest neighbor molecules that could be extracted from the crystalline structure (**Figure 3**a,b) and computed the low-energy excited states for each of them (Table S3, Supporting Information). Three molecular dimers, designated dimer 1 (D1), dimer 3 (D3), and dimer 4 (D4) mainly contribute to the excitonic properties in crystals of [3]Ph. Dimer D1 is representative of the π–π face-to-face interaction between cumulenic chains, and it lies along the *b*-axis of the crystal. Dimers D3 and D4 show a head-to-tail interaction between the chains and are oriented in the *ab*-plane of the crystal. Accordingly, D1 shows very different excitonic properties in comparison to D3 and D4, as revealed by TD-DFT calculations. For D1 the computed high- and low-energy states are at 3.17 and 2.99 eV, for D3 and D4 they are at 3.17 and 3.07 eV (Table S3, Supporting Information). Due to the molecular packing, D1 represents the prototypical H-type aggregate, with the high-energy state (3.17 eV) that is dipole active, while D3 and D4 represent J-type aggregates, with a bright, low-energy state at 3.07 eV (Figure S10, Supporting Information). Figure 3c reports the computed active electronic transition energies for the single molecule (S, black), D1 (yellow), D3 (orange) and D4 (green). Dimers D3 and D4 show a higher oscillator strength than D1, both being more intense than the single molecule case.

The low-energy absorption band observed for [3]Ph in dilute solution (2.92 eV) is thus assigned to the single molecule transition. On the other hand, the low-energy band observed in solid-state thin films (2.73 eV) corresponds to the D3- and D4-like dimers (i.e., J-type) and the high-energy band (2.97 eV) to D1-like dimers (i.e., H-type). With two types of aggregates oriented in different crystallographic directions, the low-energy band is polarized along D3 and D4, i.e., in the *ab*-plane, while the high-energy band is polarized along D1, i.e., the *b*-axis (Figure S11, Supporting Information).

Based on a molecular dimer approach, we computed the charge transfer integrals ($V_{ij}$) at the DFT level for a hole transport mechanism, and this correlates well with experimentally observed results for OFETs of [3]Ph (see Section 2.3; details in the Experimental Section). $V_{ij}$ is a measure of the electronic coupling (e.g., overlap) between the charge-carrier wavefunction of nearest neighbor sites (*i* and *j*).[66,67] Within the hopping regime, higher values of $V_{ij}$ designate higher transfer rates between hopping sites. For [3]Ph, the highest hole $V_{ij}$ is computed for D1, being 46 meV, followed by D2 and D3, both at 13 meV, while $V_{ij}$ for D4 is negligible (Table S4, Supporting Information). Such electronic couplings suggest anisotropic charge transport within the crystal, with hole percolation occurring mainly along the direction of D1 (*b*-axis) and corresponding to the π-stacking direction of the sp-chain (Figure 3a). Notably, the increased spacing along the *b*-axis, as observed from GIWAXS measurements for films deposited at 110 °C (Figure 2f and Figure S9, Supporting Information) suggests a lower electronic coupling for D1, thus indicating a less favorable platform for the charge transport under these deposition conditions. Charge transfer is also possible along the D2 (*c*-axis) and the D3 directions (*a*-axis), but with a lower probability than along D1.

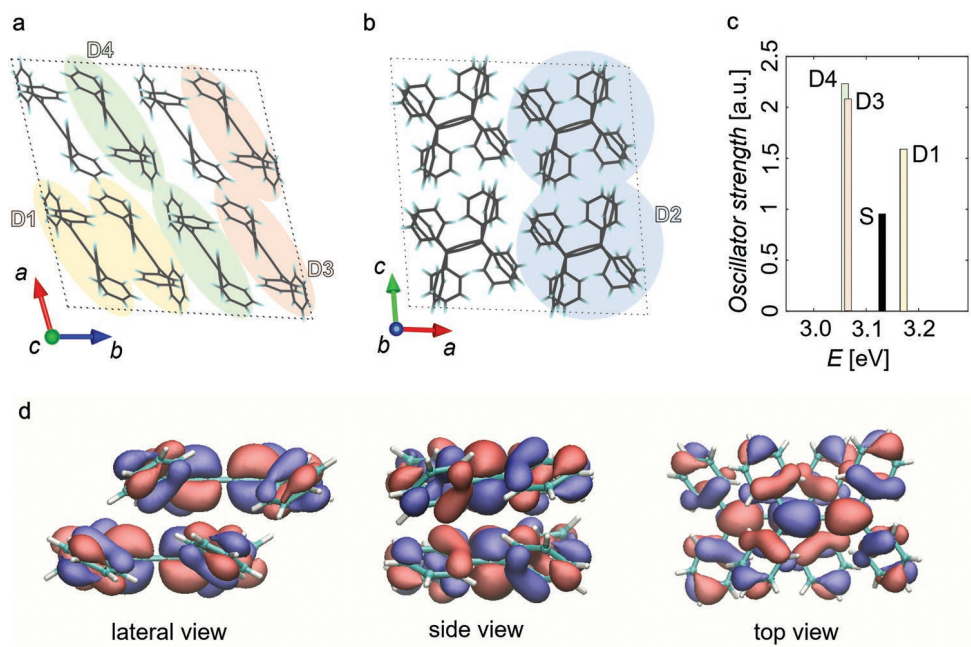

**Figure 3.** a,b) Schematic views of the molecular dimers extrapolated from the crystal structure by considering the nearest neighbor molecules in the *ab*-plane (a) and in the *ac*-plane (b). c) TD-DFT-computed allowed electronic transitions for the single molecule (S, black), dimer D1 (yellow), dimer D3 (orange) and dimer D4 (green). d) DFT-computed highest occupied molecular orbital (HOMO, isosurface 0.01 a.u.) for dimer D1.

To elucidate the importance of the sp-conjugated backbone in contributing to charge transport, the HOMO and LUMO isosurfaces for [3]Ph are reported in Figure S12a in the Supporting Information. The molecular orbitals are delocalized over the carbon chain. Furthermore, the conjugation of the sp-chain extends to the phenyl groups, which is consistent with previous evidence for cumulenic chains with varied chain lengths and end-groups.[39] The computed hole reorganization energy for [3]Ph is equal to 0.366 eV and is comparable to common organic semiconductors based on sp²-hybridized carbon scaffolds in organic electronics, such as oligothiophenes and acenes.[68,69] The major structural reorganization process occurring upon charging (i.e., adding a hole) involves the carbon chain, namely a variation of the bond length alternation (Figure S12b, Supporting Information). Additionally, Figure 3d depicts the HOMO orbital of dimer D1, which shows the highest hole transfer integral. The overlap between the orbitals of the two carbon chains can be easily observed, thus confirming the importance of the carbon backbone in determining and tuning the overall magnitude of the transfer integral for sp-systems.

## 2.3. [3]Ph-Based OFETs

Top-gate bottom-contact FETs were fabricated to investigate the electrical transport properties and ultimately evaluate the charge mobility in solution-processed [3]Ph polycrystalline thin films (**Figure 4a**). A solution of [3]Ph was deposited onto a glass substrate patterned with gold interdigitated source and drain electrodes using bar-coating at deposition temperatures ranging from 50 to 110 °C. For each deposition temperature, at least 14 devices were fabricated, with channel lengths (*L*) ranging from 2.5 to 40 μm and channel widths (*W*) from 0.2 to 2 mm. Parylene-C was selected for the dielectric layer. Finally, PEDOT:PSS gate contacts were inkjet-printed on top of the dielectric as the last step.

The characteristic transfer curves in the linear ($V_{ds} = -5$ V, black curve) and saturation ($V_{ds} = -40$ V, blue curve) regimes of a representative, optimized device measured in nitrogen atmosphere are shown in Figure 4b (films deposited at 80 °C). The curves show unipolar p-type behavior, negligible hysteresis, and ideal linear (square root) behavior of current as a function of the gate voltage in the linear (saturation) regime (Figure S13, Supporting Information). The ideality of the characteristic curves confirms robust hole mobility extraction, which is identical in the linear and saturation regime, almost independent of the gate potential ($V_g$) above threshold, and values as high as 0.11 cm$^2$ V$^{-1}$ s$^{-1}$ (Figure 4c). The mobility values extracted in the saturation regime as a function of the deposition temperature are shown in Figure 4d. The highest-performing films are those deposited at 80 °C, with an average saturation mobility of 0.08 cm$^2$ V$^{-1}$ s$^{-1}$ and a maximum value of 0.19 cm$^2$ V$^{-1}$ s$^{-1}$. This superior performance can be rationalized by considering that the 2D crystalline grains obtained from deposition at 80 °C show the lowest surface roughness and the lowest defectivity. The trend of the absorption tail of thin films as a function of the deposition temperature, as well as of the relative intensity of the absorption peaks assigned to the single-molecule and to the molecular aggregates, also support this hypothesis (Figure S1, Supporting Information). It is noteworthy that even though deposition at 110 °C leads to larger crystalline grains and fewer grain boundaries, the elongation of the crystalline structure along the *b*-axis decreases the electronic coupling, thus reducing the charge-carrier mobility. The mobility values obtained in this study represent a remarkable improvement of two orders of magnitude with respect to previous field-effect transistors based on [3]Ph.[51]

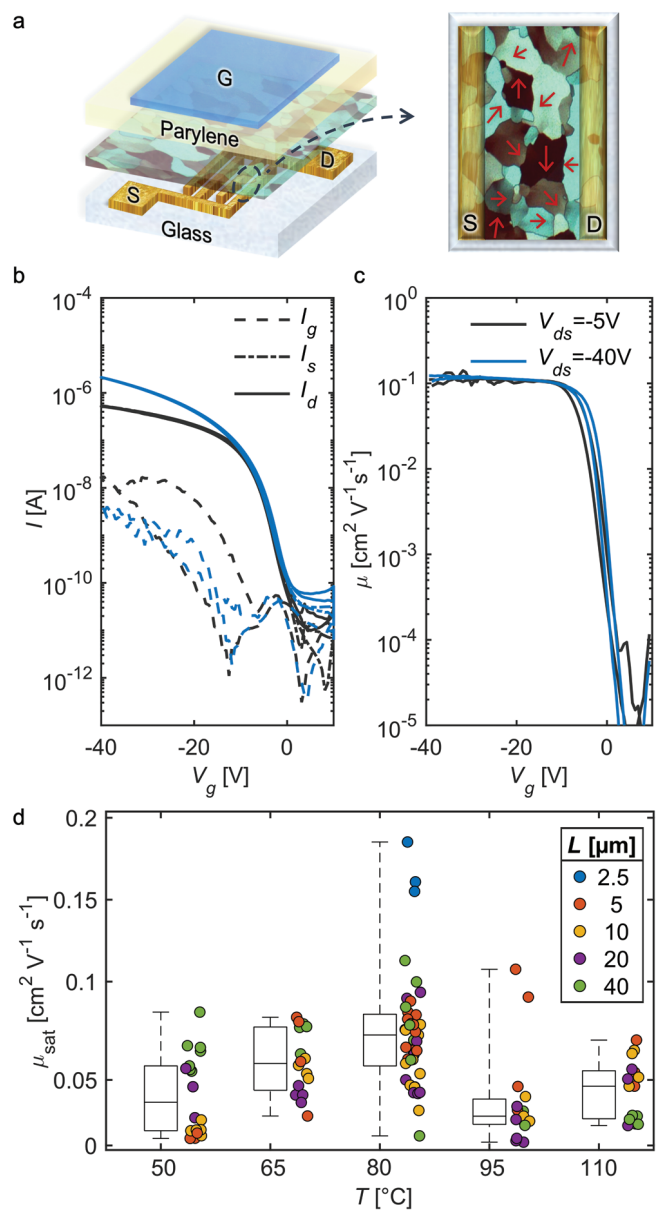

**Figure 4.** a) Structure of the OFETs fabricated with thin films of [3]Ph: source and drain electrodes are constructed from photolithographically patterned gold; a 450 nm-thick layer of Parylene-C serves as the dielectric layer ($C_{diel}$ = 5.7 nF cm⁻²) and the PEDOT:PSS gate electrode was inkjet printed. The inset represents the random azimuthal orientation of the crystalline grains within the channel area, with the red arrows oriented along the preferential transport direction in the respective domain. b) Transfer characteristic curves in the linear ($V_{ds} = -5$ V, in black) and saturation regimes ($V_{ds} = -40$ V, in blue) representative of optimized OFETs, with a channel width $W = 200$ μm and channel length $L = 40$ μm, based on a film of [3]Ph deposited at $T = 80$ °C. c) Corresponding charge mobilities as a function of the gating potential $V_g$. d) Charge mobilities extracted in saturation regime ($V_{ds} = -40$ V) as a function of the deposition temperature. At least 14 devices were tested for each deposition temperature. Each boxplot in this figure extends from the 25th to the 75th percentile of the distribution of measured mobilities, with the central bar indicating the median and the whiskers spanning from the minimum to the maximum values. The mobility values of all the transistors are reported next to each box-plot.

Given the ideality of the devices, mobility values have a reliability factor ($r$) close to 1.[70] The $I_{on}/I_{off}$, computed as the ratio between the highest and the lowest drain current values in the transfer characteristic curve, spans from $5 \times 10^4$ to $10^6$; this variability is a direct consequence of the wide range of channel lengths ($L$) that have been tested (Figure S14, Supporting Information). The complete set of parameters derived from the electrical characterization of the optimized devices deposited at 80 °C are listed in Table S5 in the Supporting Information.

We evaluated the contact resistance per unit width ($R_cW$) by the transmission-line method (TLM),[71,72] resulting in $R_cW = 41$ kΩ cm with $V_g = -40$ V (Figure S15 and Table S6, Supporting Information). The contact resistance comprises two contributions: the Schottky barrier at the metal-semiconductor interface and the access resistance.[73,74] The former originates from the energy mismatch between the HOMO level of [3]Ph (≈−5.48 eV as estimated by cyclic voltammetry, Table S1, Supporting Information) and the work function of the gold contacts (nominally at around 5.1 eV[75]). The latter is due to the vertical transport between the contacts and the active channel at the semiconductor–dielectric interface, which occurs along the *c*-axis; as previously discussed, charge transport in this crystallographic direction is less effective than in the *ab*-plane. The impact of the contact resistance on the electrical characteristics of the devices is discussed quantitatively in Figure S15b,c in the Supporting Information. The effect of contact resistance on device performance is minimized with sufficiently long channel lengths ($L > 10$ μm), at which point the channel resistance $R_{ch}$ is dominating (Figures S15 and S16, Supporting Information). Non-ideal behavior, e.g., a discrepancy in the mobility extracted in linear and saturation regimes, begins to appear in devices with $L = 2.5$ μm (Figure S14, Supporting Information). High contact resistance may explain the high threshold voltage values (on average, −8.6 V and −6.1 V in linear and saturation regime, respectively).

### 2.4. OFET Operational Stability

Since the reactivity of the sp-carbon chain represents a major concern for the application of cumulenes in electronic devices, we addressed the shelf-life stability of the [3]Ph-based OFETs. Previous differential scanning calorimetry analysis demonstrated that [3]Ph crystals are stable up to about 250 °C.[51] Therefore, the material does not thermally degrade during the deposition within the applied range of temperatures. Moreover, [3]Ph can be stored and processed in air without decomposition, when not exposed to direct sunlight. Under intense irradiation, [3]Ph is known to dimerize in the solid state.[43,76]

The decreased absorbance for films of [3]Ph during irradiation at 405 nm with a light-emitting diode (Figure S17, Supporting Information) suggests degradation, likely through a photochemical cross-linking process. Photodegradation would produce defects that are highly detrimental for charge transport, as suggested by the deteriorating transfer characteristics upon irradiation to white light (Figure S18, Supporting Information). Conversely, [3]Ph-based devices show promising operational stability when measured under dark conditions. The transfer curves in linear regime of devices measured every 10 minutes for a total time of 10 hours in nitrogen atmosphere (**Figure 5**a) and 12 hours in air (Figure 5b) have been examined. Over this timeframe, the transfer characteristics of the OFET do not degrade, and only a slight decrease in the field-effect mobilities (−2% in nitrogen atmosphere and −1% in air, Figure 5c) is observed, together with a substantial improvement of the $I_{on}/I_{off}$ ratio (+400% in nitrogen atmosphere and +3200% in air, Figure 5d).

## 3. Conclusion

This work clearly demonstrates that solution-processed thin films formed from molecules based on sp-hybridized carbon atoms can be adopted for large-area electronics. In particular, field-effect transistors based on polycrystalline thin films of a short cumulene, [3]Ph, show ideal p-type field-effect behavior and reliable hole mobility exceeding 0.1 cm$^2$ V$^{-1}$ s$^{-1}$ for optimized conditions. Furthermore, the active layer of these FETs is deposited via a simple large-area coating technique. Remarkably, non-encapsulated transistors display promising stability under dark conditions, suggesting the potential for other future applications of these molecules. We expect that the mobility values achieved in this study could be further improved. On the one hand, enhanced performance using [3]Ph is expected by considering the large anisotropy of the electronic couplings within crystals as computed by DFT/TD-DFT, the random azimuthal orientation of the crystalline grains, and the presence of grain boundaries. Enlarged crystalline grains would be beneficial due to the concurrent reduction of grain boundaries, while directional control of the crystal growth would remedy the anisotropic charge transport. On the other hand, we open the path to exploiting a vast library of cumulenic molecules, characterized by, for example, a longer conjugated backbone and tunable molecular bandgap. As well, synthetic modulation of the endcapping groups presents an exquisite opportunity to optimize crystalline packing and, thus, opto-electronic properties. The exploration of carbon atom wires in large-area organic electronics has just begun.

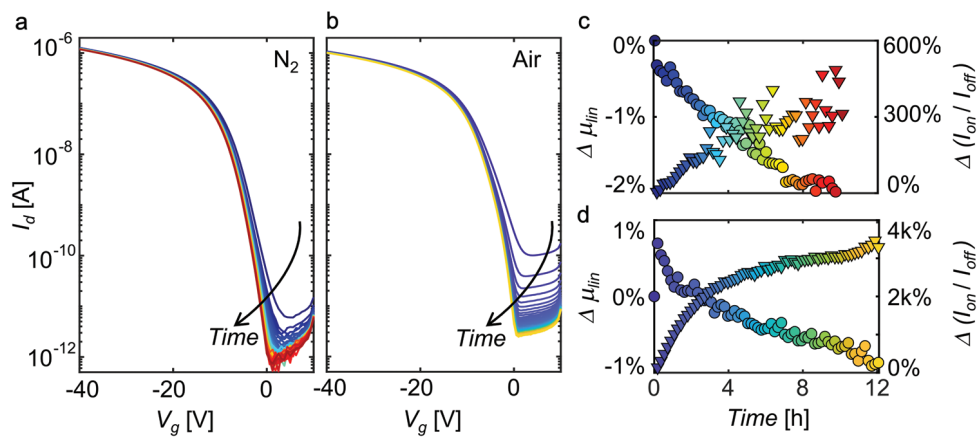

**Figure 5.** a,b) Assessment of operational stability under a nitrogen atmosphere (a) and in air (b) for OFETs with [3]Ph deposited at 80 °C ($L = 5$ μm, $W = 200$ μm). The time evolution of the transfer curves is depicted using a graded color scale (one measurement every 10 min for 1 h, $V_{ds} = -5$ V). c,d) The percentage time variations of the linear field-effect mobility, $\Delta\mu_{lin}$, and on–off current ratio, $\Delta(I_{on}/I_{off})$, related to the transfer curves measured under a nitrogen atmosphere (c) and in air (d).

## 4. Experimental Section

*[3]Ph*: Tetraphenylbutatriene ([3]Ph) has been synthesized as previously reported[77] and stored in air at about 5 °C. Single-crystal diffraction data for [3]Ph (CCDC-1817320, 1 275 120, or 1 275 121) can be obtained free of charge from The Cambridge Crystallographic Data Centre via www.ccdc.cam.ac.uk/data_request/cif.

*Deposition of Thin Films of [3]Ph*: Pure [3]Ph in powder form was dissolved in 1,2-dichlorobenzene (Sigma Aldrich) at a concentration of 10 g L$^{-1}$, stirred at 50 °C for at least 1 h, and this solution used for deposition of [3]Ph by wire-bar coating in air.

Glass substrates (low alkali 1737F Corning glasses, purchased from Präzisions Glas & Optik GmbH) were cut to a size of 2 × 2 cm, cleaned by sonication with fresh acetone and isopropyl alcohol (both purchased from Sigma Aldrich) for 10 min, and then dried under a flow of nitrogen. Subsequently, the substrates surface was cleaned and activated by an oxygen-plasma treatment for 5 min (Diener Electronic Femto Plasma, chamber pressure 40 mbar, power 100 W).

The thin films of [3]Ph were deposited under a chemical fume hood by wire-bar coating using a TQC Sheen Automatic Film Applicator AB4400, which has been equipped with a homemade system to control the gap between the substrate and the wire-bar and with a nitrogen blow system to reduce the film drying time. The employed wire-bar applicator had a wire-wound rod diameter of 6 μm. The coating speed was set to 30 mm s$^{-1}$, and the deposition temperature was controlled in the range 50–110 °C with the integrated heated bed.

*Optical Characterization*: Polarized optical microscopy images were recorded in reflection mode with a Zeiss Axio Scope A1 equipped with a single polarizer (EpiPol mode).

Absorption spectra were obtained by measuring transmission spectra on a PerkinElmer Lambda 1050 UV/Vis/NIR spectrometer.

*Local Polarized Absorption Spectra*: The UV–vis spectra in Figure S11 in the Supporting Information were recorded with a homemade confocal microscope. The incident white light firstly passed through a linear polarizer, and then it was focused on the sample through an objective with numerical aperture (NA) of 0.7 (S Plan Fluor 60×, Nikon). Successively, the beam was collected through a 0.75 NA objective (CFI Plan Apochromat VC 20×, Nikon), and finally detected by a fiber-coupled spectrometer (Avantes, AvaSpec-HS2048XL-EVO).

*Light-Induced UV–vis Evolution*: UV–vis spectra in Figure S17 in the Supporting Information were acquired by a fiber-fed spectrometer (Avantes, AvaSpec-HS2048XL-EVO) coupled with a deuterium-halogen lamp (AvaLight-D(H)-S). The photodegradation process was induced using a LED lamp (Thorlabs, M405LP1-C5) centered at 405 nm at a power of 60 mW cm$^{-2}$. This source was placed on top of the sample illuminating it uniformly.

*Films Topography and Thickness*: The surface topography of the films was imaged with a Keysight 5600LS atomic force microscope operated in tapping mode. Gwyddion software[78] was used for image processing and surface roughness calculation (details in the Supporting Information). The film thickness was measured with a KLA Tencor P-17 Surface Profiler.

*Grazing-Incidence Wide-Angle Scattering (GIWAXS)*: GIWAXS measurements were performed at the non-crystalline diffraction beamline (BL11-NCD-Sweet) at ALBA Synchrotron Radiation Facility in Barcelona (Spain). A detector (Rayonix, WAXS LX255-HS) with a resolution of 1920 × 5760 pixels was used to collect the scattering signals. Sample holder position was calibrated with chromium oxide ($Cr_2O_3$) standard. The incident energy was 12.4 eV, and the sample-to-detector distance was set at 216.5 mm. The angle of incidence $\alpha_i$ was 0.11° and the exposure time was 1 s. 2D-GIWAXS patterns were corrected as a function of the components of the scattering vector with a MATLAB script developed by Aurora Nogales and Edgar Gutiérrez.[79] Thin films were cast onto highly doped silicon substrates following same processing route used for the device fabrication.

Miller indexing and simulation of the lattice unit were performed with the MATLAB-SIIRkit package, designed and coded by Savikhin et al.;[80] using the diffraction peaks localized in the experimental GIWAXS maps as input data. The patterns were previously transformed into q-space and corrected by taking into account the curvature of the Ewald sphere (missing wedge). The presented results are the best fittings that adjust to all the experimental diffraction peaks.

*OFETs Fabrication*: [3]Ph-based OFETs were fabricated with a bottom-contact top-gate architecture. Bottom source and drain interdigitated electrodes were defined by standard photolithography onto the glass substrates and deposited by thermal evaporation of 30 nm-thick Au with a 3 nm-thick Cr adhesion layer. The [3]Ph active layer was deposited by wire-bar coating according to the procedure described above. About 500 nm-thick dielectric layer of poly(chloro-p-xylene)-C (Parylene-C, dimer purchased from Specialty Coating Systems) was deposited by CVD with a SCS Labcoater 2—PDS2010 system. PEDOT:PSS (Clevios PJ700 formulation, purchased from Heraeus) gate contacts were inkjet-printed on top of the dielectric layer by means of a Fujifilm Dimatix DMP2831.

*Electrical Characterization*: OFETs transfer and output electrical characteristics were measured with a semiconductor parameter analyzer (Agilent B1500A) in a nitrogen glove box on a Wentworth Laboratories probe station. The transfer curves for the stability test in Figure 5b were measured in air with the same setup.

*DFT/TD-DFT Calculations on Single Molecule and Dimers*: Single-molecule and dimer DFT/TDDFT calculations were performed by using the $\omega$B97X-D3 range-separated functional and the double (triple) split valence basis set with 6–31G* (6-311G*) polarization function. Calculations were carried out with the Gaussian16/C.01 code.[81] Electronic transfer integrals were computed using the symmetry projection dimer methods as reported in ref. [82]. Exciton couplings were evaluated by performing TD-DFT calculation on the dimers and by using the dimer-splitting method. The computed hole reorganization energy of [3]Ph was evaluated via the adiabatic potential energy method at the DFT level ($\omega$B97X-D3/cc-pVTZ).

## Supporting Information

Supporting Information is available from the Wiley Online Library or from the author.

## Acknowledgements

E.G.F. acknowledges the support through the EU Horizon 2020 research and innovation program, H2020-FETOPEN-01-2018-2020 (FET-Open Challenging Current Thinking), "LION-HEARTED", grant agreement no. 828984. C.S.C. acknowledges funding from the European Research Council (ERC) under the European Union's Horizon 2020 research and innovation program ERC-Consolidator Grant (ERC CoG 2016 EspLORE grant agreement no. 724610, website: www.esplore.polimi.it). R.R.T. acknowledges funding from the Natural Sciences and Engineering Research Council of Canada (NSERC) and the Canada Foundation for Innovation (CFI). This work was partially supported by the European Union's H2020-EU.4.b. – Twinning of research institutions "GREENELIT", grant agreement number 951747. GIWAXS experiments were performed at BL11 NCD-SWEET beamline at ALBA Synchrotron (Spain) with the collaboration of ALBA staff. This work was in part carried out at Polifab, the micro- and nanotechnology centre of the Politecnico di Milano.

Open access funding provided by Istituto Italiano di Tecnologia within the CRUI-CARE Agreement.

## Conflict of Interest

The authors declare no conflict of interest.

## Data Availability Statement

The data that support the findings of this study are available from the corresponding author upon reasonable request.